\newcommand{\PaperTitle}{ECSeptional DNS Data: Evaluating Nameserver ECS Deployments with Response-Aware Scanning}

\documentclass[acmsmall]{acmart}

\setcopyright{cc}
\setcctype{by}
\acmJournal{PACMNET}
\acmYear{2025} \acmVolume{3} \acmNumber{CoNEXT2} \acmArticle{11} \acmMonth{6} \acmPrice{}\acmDOI{10.1145/3730977}

\usepackage{packages}
\usepackage{textpos}

\acmJournal{PACMNET}

\makeatletter
\newcommand\EatSpacesHack{\@bsphack\@esphack}
\makeatother
\iffalse{}
\newcommand\reviewfix[1]{{\color{red}\sffamily\bfseries [RF:\ #1]}\EatSpacesHack}
\else
\newcommand\reviewfix[1]{\EatSpacesHack}
\fi
\makeatother

\newcommand*{\dittostraight}{---\textquotedbl---}

\newcommand{\jz}[1]{\hl{\textbf{JZ: #1}}}
\newcommand{\ps}[1]{\hl{\textbf{PS: #1}}}
\newcommand{\kv}[1]{\hl{\textbf{KV: #1}}}
\newcommand{\mj}[1]{\hl{\textbf{MJ: #1}}}
\newcommand{\fh}[1]{\hl{\textbf{FH: #1}}}
\newcommand{\og}[1]{\hl{\textbf{OG: #1}}}

\renewcommand{\jz}[1]{\iffalse\hl{\textbf{JZ: #1}}\fi}
\renewcommand{\ps}[1]{\iffalse\hl{\textbf{PS: #1}}\fi}
\renewcommand{\kv}[1]{\iffalse\hl{\textbf{KV: #1}}\fi}
\renewcommand{\mj}[1]{\iffalse\hl{\textbf{MJ: #1}}\fi}
\renewcommand{\fh}[1]{\iffalse\hl{\textbf{FH: #1}}\fi}
\renewcommand{\og}[1]{\iffalse\hl{\textbf{OG: #1}}\fi}

\begin{document}

\setlength{\TPHorizModule}{\paperwidth}
\setlength{\TPVertModule}{\paperheight}
\begin{textblock}{0.802}(-0.025,-0.09)
  \fbox{\begin{minipage}{\linewidth}
    \noindent
    \footnotesize
    If you cite this paper, please use the ACM CoNEXT reference: Patrick Sattler, Johannes Zirngibl, Fahad Hilal, Oliver Gasser, Kevin Vermeulen, Georg Carle, and Mattijs Jonker. 2025. ECSeptional DNS Data: Evaluating Nameserver ECS Deployments with Response-Aware Scanning. Proc. ACM Netw. 3, CoNEXT2, Article 11 (June 2025), 25 pages. https://doi.org/10.1145/3730977
  \end{minipage}}
\end{textblock}

\title[ECSeptional DNS Data]{\PaperTitle}

\author{Patrick Sattler}
\email{sattler@net.in.tum.de}
\orcid{0000-0001-9375-3113}
\affiliation{%
  \institution{Technical University of Munich}
  \city{Munich}
  \country{Germany}
}
\author{Johannes Zirngibl}
\email{jzirngib@mpi-inf.mpg.de}
\orcid{0000-0002-2918-016X}
\affiliation{%
  \institution{Max Planck Institute for Informatics}
  \city{Saarbrücken}
  \country{Germany}
}

\author{Fahad Hilal}
\email{fhilal@mpi-inf.mpg.de}
\orcid{0009-0008-8123-4496}
\affiliation{%
  \institution{Max Planck Institute for Informatics}
  \city{Saarbrücken}
  \country{Germany}
}

\author{Oliver Gasser}
\email{oliver@ipinfo.io}
\orcid{0000-0002-3425-9331}
\affiliation{%
  \institution{IPinfo}
  \city{Seattle}
  \country{United States}
}

\author{Kevin Vermeulen}
\email{kevin.vermeulen@laas.fr}
 \orcid{0000-0001-6168-7887}
\affiliation{%
  \institution{LIX, CNRS, Ecole Polytechnique}
  \city{Palaiseau}
  \country{France}
}

\author{Georg Carle}
\email{carle@net.in.tum.de}
\orcid{0000-0002-2347-1839}
\affiliation{%
  \institution{Technical University of Munich}
  \city{Munich}
  \country{Germany}
}

\author{Mattijs Jonker}
\email{m.jonker@utwente.nl}
 \orcid{0000-0001-5174-9140}
\affiliation{%
  \institution{University of Twente}
  \city{Enschede}
  \country{The Netherlands}
}
\renewcommand{\shortauthors}{Patrick Sattler et al.}

\begin{abstract}
    DNS is one of the cornerstones of the Internet.
    Nowadays, a substantial fraction of DNS queries are handled by public resolvers (e.g., Google Public DNS and Cisco's OpenDNS) rather than ISP nameservers.
    This behavior makes it difficult for authoritative nameservers to provide answers based on the requesting resolver.
    The impact is especially important for entities that make client origin inferences to perform DNS-based load balancing (\eg CDNS).
    The EDNS0 Client Subnet (ECS) option adds the client's IP prefix to DNS queries, which allows authoritative nameservers to provide prefix-based responses.
    Previous work showed the potential of data collected during ECS scans.
    Infrastructure can be uncovered, and operators' subnet-specific behavior can be observed.

    In this study, we introduce a new method for conducting ECS scans. Our method significantly reduces the required number of queries by up to \sperc{97} compared to state-of-the-art techniques and allows us to provide new insights into ECS behavior.
    Our approach is also the first to facilitate ECS scans for IPv6.
    Due to its vast address space, we have developed and analyzed different IPv6 scanning approaches.
    We conduct a comprehensive evaluation of the ECS landscape, examining the usage and implementation of ECS across various services.
    Overall, \sperc{53} of all nameservers support prefix-based responses.
    Furthermore, we find that Google nameservers do not comply with the Google Public DNS guidelines.
    Additionally, we observe that certain operators (\eg AWS Route53) exclusively employ a single specific scope prefix length without aggregation, potentially affecting resolver cache efficiency.
    Lastly, we make our tool and data publicly available to foster further research in the area.

\end{abstract}



\maketitle

\section{Introduction}

Popular services on the Internet are commonly served by multiple CDN edge servers.
CDNs offer load distribution and low latency services to ensure user satisfaction and improve conversion rates~\cite{arapakis2014websearchperformance,bernardi2019booking,qiu2020slo}.
Therefore, service operators aim to optimize the load distribution.

Two well-known techniques for operators to direct the initial connection of a user to a nearby vantage point are IP anycast and DNS load balancing.
This work evaluates a particular aspect of DNS-based load balancing: the \ecs option~\cite{rfc7871}.
The \ecs option is a solution to the hidden client problem for authoritative nameservers.
Without it, the authoritative nameserver only knows the recursive resolver IP address, which originated the DNS query.
Traditionally, it could only use that recursive resolver IP address to perform topology- or geolocation-based load balancing.
With \ecs{}, the resolver includes the client's subnet in the query.
Therefore, authoritative nameservers can use this information to provide tailored responses to the client.

Major cloud providers such as Google~\cite{googleecs}, Amazon~\cite{amazonecs}, Akamai~\cite{akamaiecs}, and Cloudflare~\cite{cloudflareecs} all provide \ecs query responses and offer it as a product to their customers.
On the software side, \ecs is also widely adopted, \eg BIND~\cite{bindecs}, Knot DNS~\cite{knotecs}, and PowerDNS~\cite{powerdnsecs} support scoped responses using \ecs{}.
Hence, \ecs is available to a large customer base.
Additionally, prominent public recursive resolvers such as Google Public DNS and Cisco's OpenDNS also support \ecs{}. %
The importance of recursive resolvers supporting \ecs is well-documented~\cite{calder2019ecs}.
Existing work has confirmed ECS's effectiveness in reducing client latency~\cite{otto2012ecs,sanchez2013ecs,warrior2017drongo}.
Others~\cite{streibelt-ecs,calder2013ecs,otto2012ecs} have looked at ECS support on the authoritative DNS infrastructure of top-list domains.

Using \ecs{} requires major effort by providers to effectively map clients to \acp{pop}.
Determining how they use \ecs{}, their mapping strategy, and whether they conform to the standard is important.
This knowledge can help resolvers, network operators, and clients to better understand the network and perceived quality of experience.
The information on how providers map clients to \acp{pop} can be extracted using \ecs scans.
The scan results can be used to better understand the provider's infrastructure and serve as a data source for various use cases.

This paper presents a novel approach to perform IPv4 and IPv6 ECS scans.
Unlike existing approaches, we keep the state from responses already received.
By doing so, we are able to significantly reduce the number of required queries, alleviating the load on third-party infrastructure.
We reveal and investigate previously unseen load balancing behaviors.
Google uses daily patterns to update its client-server mapping.
Moreover, we leverage our efficiency-increasing approach to revisit \ecs support among popular domain names a decade after prior work~\cite{streibelt-ecs,calder2013ecs,otto2012ecs} and for a more diverse set of top lists.
Our results show a significant increase in \ecs adoption among popular domains:
\sperc{79} signal support for \ecs{}, and \sperc{40} also provide subnet-specific responses.

\vspace{0.4em}
\noindent{}The main contributions of this work are:

\begin{itemize}[topsep=0pt]
    \item[\first] We present a novel \ecs scanning approach, the first that supports \ecs scans for IPv6 (see \Cref{sec:approach}). \reviewfix{2} Due to its response awareness, our approach significantly reduces the number of queries to cover the address space. Additionally, it provides a fine-grained configuration of query limits per prefix length.
    \item[\second] We use our scanning approach to evaluate the \ecs{} behavior of nine popular domains served by four providers (see \Cref{sec:response-aware}).
    Our approach reduces the number of queries needed for the IPv4 address space by up to \sperc{97} over the state of the art.
    Based on our analysis, the best set to uncover IPv6 infrastructure is a combination of BGP and IP geofeed prefixes.
    \item[\third] We revisit and explore the prevalence of \ecs among top-listed domains and provide insights into different nameserver operator behaviors (see \Cref{sec:ecslandscape}). We find that Cloudflare provides \ecs load balancing as a service for their customers but also responds with scoped answers for their own anycast addresses.
    \item[\fourth] We perform a detailed analysis of \ecs properties for Google, Meta, and AWS Route53 authoritative nameservers (see \Cref{sec:nsproperties}).
    We uncover RFC-violating behavior by Google, which its own public DNS resolver does not accept.
    \item[\fifth] We publish the scanner's source code~\cite{ecsplorer} and make our scanning data publicly available~\cite{datatum}.
\end{itemize}

\section{EDNS0 Client Subnet Option}
\label{sec:ecsbackgound}

The separation of DNS into clients, recursive resolvers, and authoritative nameservers results in a fundamental problem if DNS is used for (geo-)load balancing:
The authoritative nameserver is aware of the resolver and its IP address (thus location).
However, the resolver does not have to be geographically or topologically close to the client (\eg{} using public resolvers).

\begin{figure}
    \includegraphics{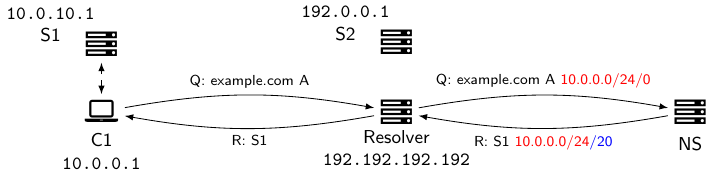}
    \caption{Example ECS flow: The client C1 is topologically close to the server S1 but uses the resolver topologically close to S2. With the client subnet information in the query (IP address 10.0.0.0 with source prefix length 24), the nameserver can select the correct response and send it back with the original source prefix length 24 and the scope prefix length of 20. The resolver can thus cache the response for all clients within 10.0.0.0/20.}
    \label{fig:ecs_background}
\end{figure}

To overcome this problem, RFC7871~\cite{rfc7871} describes \ecs{} as an \ac{edns}~\cite{rfc6891} option.
This option allows a resolver to add a client's subnet to a query.
Using the client's subnet instead of its IP address preserves privacy, reduces the load on nameservers, and allows for appropriate caching.
Authoritative nameservers can use this subnet information to tailor responses to the client.
\Cref{fig:ecs_background} shows the general functionality and information flow of \ecs{}.

The \ecs option consists of four properties: The IANA address family~\cite{ianaaddressfamily}; an address; and a source and scope prefix length.
While the address and its family are self-explanatory, we delve deeper into the properties of the prefix length.
The resolver selects the source prefix length based on its ability to cache responses without resource exhaustion.
RFC7871 suggests a maximum value of 24 for IPv4 and 56 for IPv6 to preserve user privacy.

The authoritative nameserver uses the client information to compose its response.
It sets the so-called scope length to the prefix length for which the answer is valid.
The resolver must cache the response accordingly.
A scope prefix length less specific or equal to the source prefix length indicates that the response is valid for this (less specific) prefix.
A scope prefix length of 0 covers the complete address space.
If the scope prefix length is more specific than the source prefix length, the authoritative nameserver indicates that the provided prefix length is not specific enough.
The resolver can cache based on its capabilities (the initially sent source prefix length) or rerun the query with a more specific source prefix length.
In \Cref{fig:ecs_background}, the resolver chooses a source prefix length of 24, while the nameserver responds with a less specific scope prefix length of 20.

\noindent
\textbf{Terminology:}
In the remainder of this paper, we use the term \emph{scoped response} to indicate a response that contains an \ecs scope larger than zero.
This response can only be cached for the prefix within the \ecs{} option.
We identify nameservers and domains exhibiting such behavior as \ecs \emph{enabled}/\emph{supporting}.
They are \emph{using} \ecs iff they have \ecs \emph{enabled} \emph{and} if different \acp{rrset} are returned for distinct \ecs prefixes.

\section{Related Work}
\label{sec:relatedwork}

We identify three directions relevant to this research: The latency reduction achieved through \ecs, previous active \ecs scanning approaches, and the research use cases.

\noindent\textbf{ECS Latency Reduction:}
Several research groups~\cite{otto2012ecs,streibelt-ecs,calder2013ecs,sanchez2013ecs} showed the potential of \ecs to reduce end-to-end latency.
In 2012, \citeauthor{otto2012ecs}~\cite{otto2012ecs} performed an evaluation of \ecs and found that \sperc{9} of Alexa top 1k sites used the \ecs option.
They demonstrated that using the \ecs option substantially benefited the client's end-to-end latency.
The median improvement for services with Akamai was \sperc{40}.
In 2013, \citeauthor{sanchez2013ecs}~\cite{sanchez2013ecs} confirmed these results with \sk{90} users on their measurement platform.
They showed that users from Oceania benefited the most from using \ecs.
More recent research by \citeauthor{calder2019ecs}~\cite{calder2019ecs} in 2019 evaluated the adoption of \ecs at local resolvers by analyzing authoritative nameserver logs from the Azure cloud.
\sperc{40} of observed queries included the \ecs option.
These results show that services benefit from deploying \ecs{} and that resolvers support it.
For large providers, this benefit outweighs the client-server mapping complexity introduced by \ecs{}.
As a result of this previous research, we expect to find \ecs{} deployments within top lists.

\noindent\textbf{ECS Scanning:}
\citeauthor{streibelt-ecs}~\cite{streibelt-ecs} and \citeauthor{calder2013ecs}~\cite{calder2013ecs} performed active IPv4 \ecs scans to uncover service infrastructure in 2013.
\citeauthor{streibelt-ecs} found that \sperc{13} of Alexa domains seemed to enable the \ecs option, but only \sperc{3} actually used it.
Nevertheless, they showed the importance of \ecs by analyzing an \ac{isp} network with more than \sk{10} end-users.
In total, \sperc{30} of traffic volume involves services using \ecs{}.
They used different prefix lists, including BGP dumps from RIPE RIS~\cite{riperis} and Routeviews~\cite{routeviews} as target prefixes for their \ecs scans.

\citeauthor{calder2013ecs}~\cite{calder2013ecs} collected \ecs scan data over ten months to map Google's serving infrastructure.
Their scans to map Google's serving infrastructure used all routable /24 prefixes from Routeviews.
According to their results, \ecs scans found around \sperc{20} more addresses than those collected from open resolvers.
Moreover, their long-running scan shows that a single scan does not reveal all addresses, and only an accumulated dataset can provide a more complete view.
\citeauthor{calder2013ecs} use Google's own resolver as an intermediary to forward the \ecs queries.

\reviewfix{1}
\citeauthor{kountouras2021understanding}~\cite{kountouras2021understanding} evaluated the use of \ecs at authoritative nameservers.
They see the continuous growth of \ecs usage in general and also for sites that do not profit from \ecs{}.
\citeauthor{kountouras2021understanding} conduct a small-scale active measurement by issuing queries to recursive resolvers, which add \ecs information to the iterative queries towards authoritative nameservers.
This scan aimed to determine the support of ECS by domains using a single query.
They determined that the support across the Alexa Top 1M domains increased from \sk{161} in 2015 to \sk{418} in 2019.
They did not provide data on nameserver adoption.

Related to these efforts, we revisit and substantially extend results for a diverse set of top lists in this paper.
We devise and publish a significantly better-performing scanning approach for IPv4 (see \Cref{sec:response-aware}).
To our knowledge, our \ecs scanner is the first public tool specifically dedicated to perform \ecs scans.
Moreover, we introduce support for structural IPv6 \ecs scans for the first time.
We perform scans for popular domains and compare our results to those of previous work.
Additionally, we perform infrastructure coverage validation scans using RIPE Atlas.

\noindent\textbf{Research Use Cases:}
\ecs also enables researchers to learn more about service deployments.
\citeauthor{jiang2021identifyingclients}~\cite{jiang2021identifyingclients} use \ecs queries to resolvers to determine if subnets contain end users.
They validate their results using Microsoft CDN ground truth data.
\citeauthor{streibelt-ecs}~\cite{streibelt-ecs} and \citeauthor{calder2013ecs}~\cite{calder2013ecs} map the responses to users and cluster them to better understand the service's load balancing behavior.
\ecs scans can also be used to uncover IPv6 offnets, as presented by \citeauthor{hilal2024ipv6hypergiants}~\cite{hilal2024ipv6hypergiants}.
\citeauthor{sattler2022privaterelay}~\cite{sattler2022privaterelay} use full address space \ecs scans to obtain the ingress nodes of iCloud Private Relay~\cite{privaterelay}.

\citeauthor{calder2013ecs} developed the \emph{client-centric front-end geolocation (CCG)} approach to localize the front-end infrastructure via \ecs scanning.
It uses geolocation databases to localize the client prefixes that trigger the responses.
In a follow-up work, \citeauthor{fan2015useraffinity}~\cite{fan2015useraffinity} measured the affinity of users to front-end clusters by evaluating \ecs scan data.
By applying the \emph{CCG} approach, they find that users are remapped to different clusters, often more than \SI{1000}{\kilo\meter} apart.
\citeauthor{warrior2017drongo}~\cite{warrior2017drongo} presented their client-side approach to find the service's front-end with the lowest latency based on \ecs.
They and others~\cite{aldalky2019ecs} bring up the issue of operators (\eg Akamai) restricting \ecs access to pre-approved recursive resolvers. %
Their evaluation and interpretation show that unrestricted \ecs adoption is substantially more beneficial compared to its drawbacks.

This list of use cases shows the importance of \ecs scanning to research.
Our dedicated scanner can help to promote new research due to its ease of use.
Our approach offers a more efficient and ethical foundation for future research that relies on \ecs scans.
In this work, we explore the prevalence of restricted \ecs in top lists and perform scans towards Google's open resolvers and authoritative nameservers in \Cref{ssec:pubresolver}.
While it is not the primary focus of this work, we shortly highlight potential use cases in \Cref{sec:showcases}.

\section{Response-Aware ECS Scanner}
\label{sec:approach}

Previous work indicates significant support for \ecs at nameservers.
To better understand the ECS ecosystem, an efficient scanning technique is needed.
Existing tools that can be used to send \ecs queries, including dig~\cite{bind} and ZDNS~\cite{zdns}, are either not scalable (dig), or only allow static prefixes to be sent (dig and ZDNS). \reviewfix{2}
Therefore, we created a dedicated \ecs scanner called \emph{ECSplorer}~\cite{ecsplorer}.
Our scanner is a highly configurable stateful tool that also supports IPv6.
We aimed to solve two concrete goals with our dedicated \ecs scanner:
\begin{itemize}
    \item[\first{}] Uncovering the service's infrastructure by collecting all available IP addresses.
    \item[\second{}] Understanding \ecs load balancing behavior.
\end{itemize}
While the first goal aims to use a small set of queries to find all possible addresses, the second seeks detailed information on all distinct \ecs scopes the authoritative nameserver uses.
\Cref{sec:showcases} includes preliminary evaluations showing future potential for some \ecs research use cases.
We additionally aim for an efficient and ethical scanning approach to reach these goals.
Our second goal, in particular, requires sending many queries.
Therefore, we take extra caution to not perform unnecessary queries to lower the load on nameservers.

\noindent\textbf{IPv4 Approach:}
Similar to related work~\cite{calder2013ecs,streibelt-ecs}, we iterate over the BGP-announced address space and perform queries with all relevant subnets.
We obtain the BGP-announced address space by extracting all announced prefixes from BGP dumps (\eg from RouteViews~\cite{routeviews}).
Our scanner supports arbitrary source prefix lengths.
In this work, we choose a source prefix length of 24 bits.
In contrast to previous work, our approach does not need to scan all announced /24 IPv4 prefixes.
Instead, it respects the scope prefix length by the authoritative nameserver and skips the covered address space (\eg in \Cref{fig:ecs_background} we would stop querying the full /20).

Moreover, a per-prefix length threshold can be provided to limit the number of queries for each prefix of the configured length (see \Cref{sec:appendixquerylimits}).
Our scanner does not increase the specificity of the source prefix length over the configurable value.
Thus, more specific scoped responses (\eg a scope of 28 in \Cref{fig:ecs_background}) would not trigger more specific queries when the source prefix length is set to 24.
The scanner can also be configured to perform queries for a limited number of prefixes in unrouted and RFC1918~\cite{rfc1918} address space.

In order to efficiently implement this approach, we build a Patricia trie~\cite{knuth1973art} mapping out the BGP-announced address space.
Each node contains all relevant information for the scanner (\eg number of scans).
This way, the scanner can walk through the address space to compute recursively if any prefix threshold has been reached (see \Cref{sec:appendixquerylimits}).
We use Go and its parallelization features to implement our approach efficiently.

To overcome nameserver misbehavior, such as eventual /0 scope prefix lengths, we limit the accepted scope prefix length to a minimum of /8, \ie the maximum assigned prefix length from RIRs to \acp{as}.
Therefore, we ignore answers with a scope prefix length less specific than /8 when deciding which subnet to scan next. %
This property also enables us to parallelize domains per /8, further increasing the scan performance.
In reality, we never had to parallelize a single domain with more than four /8 scans to stay within our query limits (see \Cref{sec:appendixquerylimits}).
Our implementation runs on low-end hardware, and we also ran it from a virtual machine with 2 cores and 2GB of RAM without any issues.

\noindent\textbf{IPv6 Approach:}
To account for the vast IPv6 address space, we extend our scanning approach with IPv6-tailored features.
We aim to cover as much of the BGP-announced address space as possible and in as much detail as possible.
While IPv4 has approximately \sm{12} routable /24 prefixes, IPv6 has 15 billion /48 prefixes.  %
RFC7871~\cite{rfc7871} even suggests using a maximum of 56 for the scope prefix length, resulting in \num{3.8} trillion possible prefixes.
Sending over a thousand times more queries compared to IPv4 is neither reasonable nor ethical.
Hence, we choose a source prefix length of 48 and seed our scan with a predefined list of prefixes to limit the address space.
This seed list limits the address space to scan for, and the scanner will select random subnets inside this space.
The scanner is still response-aware and does not issue queries for subnets that are inside a returned scope. %
The scan completes when there is no more seeded address space to be scanned within the defined limits.

We also added an option to scan every prefix from the seed list at least once, \eg{} to scan all BGP-announced prefixes.
This option ensures that at least one query is sent for each announced prefix, even if prefix limits have already been reached.
Other possible seed list sources are the \acp{irr} or the IPv6 Hitlist~\cite{gasser2018clusters, zirngibl2022rustyclusters}.
We evaluate and compare the lists in \Cref{ssec:resp-aware-v6}.
Like regular port scanning in IPv6, selecting the target list is important for IPv6 \ecs{} scans (see \Cref{sec:response-aware}).

\reviewfix{2}
Our scanner is also capable of only issuing queries for a predefined list of \ecs subnet values.
This feature can be especially useful to scan a limited set of IPv6 subnets that are deemed to be representative of the address space that will be analyzed.
A static input list can be used to obtain comparable results over time.

\noindent\textbf{Ethics:}
At the core of our approach are ethical considerations to reduce the number of queries while obtaining the same information.
We do not process any user data, and we apply strict rate limiting to our active scans.
A detailed description of our measures can be found in \Cref{sec:ethics}.

\section{Evaluation of Response-Aware Scanning Technique}%
\label{sec:response-aware}

In this section, we analyze the impact of our approach on the number of issued queries and observed responses.
We select nine top-listed domains to perform full IPv4 address space scans (no limits applied) using our stateful scanning approach.
Additionally, we evaluate different IPv6 address space scanning approaches and compare our implementation with existing techniques.
Finally, we evaluate the returned scope prefixes to gain insights into different rates of query savings.

\begin{table}
    \caption{Overview of IPv4 response-aware address space scans. An \acs{rrset} represents the combination of resource records within each answer. The \ecs scopes column contains the observed number of distinct \ecs response scope lengths. \url{tiktok.com} is not hosted on AWS, but uses Route53 for DNS services.}%
    \label{tbl:fullscanstats}
    \centering
    \footnotesize
    \begin{tabular}{lrrrrc}
        \toprule
        & Queries & \acsp{rrset} & Addresses & \ecs Scopes & NS Operator \\
        \midrule
        en.wikipedia.org   & \sm{0.4} & 5 & 5 & 26 & Wikimedia \\
        \midrule
        m.facebook.com     & \sm{1.2} & 139 & 139 & 17 & \multirow{3}{*}{Meta}\\
        web.whatsapp.com   & \sm{1.2} & 139 & 139 & 17 & \\
        www.instagram.com  & \sm{1.2} & 139 & 139 & 17 & \\
        \midrule
        www.google.com     & \sm{1.7} & \num{32531} & 2195 & 26 & \multirow{2}{*}{Google} \\
        www.youtube.com    & \sm{1.7} & \num{322534} & 1839 & 26 &\\
        \midrule
        tiktok.com         & \sm{11.8} & \num{3687} & 636 & 1 & \multirow{3}{*}{AWS Route53}\\
        www.amazon.com     & \sm{11.8} & 323 & 323 & 1 & \\
        www.primevideo.com & \sm{11.8} & 161 & 161 & 1 &\\
        \bottomrule
    \end{tabular}
\end{table}

\begin{figure*}
    \begin{subfigure}{.347\textwidth}
        \centering
        \includegraphics{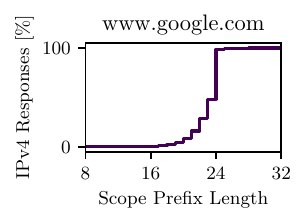}
    \end{subfigure}
    \begin{subfigure}{.315\textwidth}
        \centering
        \includegraphics{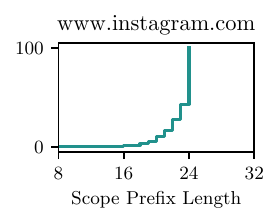}
    \end{subfigure}
    \begin{subfigure}{.315\textwidth}
        \centering
        \includegraphics{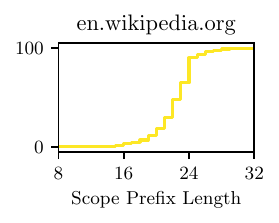}
    \end{subfigure}
    \caption{Scope prefix lengths from our IPv4 response-aware scans. AWS always uses /24. See \Cref{fig:scopesv6} for IPv6.}
    \label{fig:scopes}
\end{figure*}

\vspace*{-0.5em}
\subsection{IPv4 Results}

For our evaluation, we analyze a scan from March 20, 2024 towards the authoritative nameservers of nine popular domains supporting \ecs (see \Cref{tbl:fullscanstats}).
The nameservers of these domains are provided by Google, Meta, Amazon Web Services (AWS), and Wikimedia.
While the domains associated with Google, Meta, and Wikimedia are served by nameservers of the sites own organization, AWS's Route53 also provides DNS service for \url{tiktok.com}.
Not only TikTok but any AWS customer can use Route53 to provide subnet-scoped \ecs responses~\cite{amazonecs}.
Cloudflare's customers can use a similar feature when using Cloudflare load balancers~\cite{cloudflareecs}.

\noindent\textbf{Scan Volume Analysis:}
\Cref{tbl:fullscanstats} shows the number of issued queries per domain.
We can clearly see the uniform behavior of nameservers for each operator.
Our approach needs nearly the same number of queries for all domains hosted by the same operator, even though \acp{rrset} and uncovered addresses differ.
Our scanner issues the fewest queries (\sm{0.4}) towards Wikimedia's nameservers.
According to Wikimedia's documentation~\cite{wikipediadnsgeo}, the MaxMind geolocation database~\cite{maxmind} and the gDNS geoip plugin~\cite{gdns} are used to deploy \ecs{}.
This relatively low number of queries is confirmed by Wikimedia only using six different sites~\cite{wikipediadcs} at the time of writing, for which less fine-grained results are needed.
While we see all six Wikimedia data centers with other measurements, the Carrollton, Texas one was out of service~\cite{wikipediaswitchover} during this scan.

With Meta and Google, our scanner sends \sm{1.2} and \sm{1.7} queries to cover the full address space.
While Meta responds with a single resource record per response and provides exactly 139 addresses for each domain, Google does respond with multiple resource records and has different address mappings for Google and YouTube.
Conversely, AWS nameservers always respond with a scope prefix length of 24.
Therefore, our approach is forced to scan each announced /24 prefix while ignoring IANA special-purpose~\cite{ianaspecialpurpose} and non-announced prefixes.

TikTok always returns four records per \ac{rrset}, and we observe \num{3687} \acp{rrset}.
Google and YouTube exhibit a substantially larger number of \acp{rrset} than the number of distinct addresses found (\cf \Cref{tbl:fullscanstats}).
While Google uses one or six address records in its responses, we observe answers with everything between one and 16 records for YouTube.
Therefore, the number of possible combinations leads to this large number of observed \acp{rrset}.
Our second goal is to understand the load balancing strategy of nameservers.
The number of combinations hints at the complexity of the mapping used by the nameserver.
While all other operators stick to a 1:1 mapping between vantage points and client subnets, Google, YouTube, and TikTok use a 1:n mapping.

\noindent\textbf{Query Savings:}
As AWS is always responding with a scope prefix length of 24, we can use AWS's query volume to determine the query savings our response-aware approach eliminated in comparison to scanning all /24 prefixes as done by \citeauthor{calder2013ecs}~\cite{calder2013ecs} or \citeauthor{streibelt-ecs}~\cite{streibelt-ecs}.
Our approach eliminates \sperc{97} of queries for Wikimedia, \sperc{90} for Meta, and \sperc{86} for Google.
These numbers demonstrate the importance of using a response-aware scanner to eliminate a large part of superfluous DNS queries.

Next, we analyze the different scope prefix lengths returned for Google, Instagram (Meta), and Wikipedia (Wikimedia).
As shown in \Cref{fig:scopes}, Wikimedia returns more answers with lower scope prefix lengths than Google and Meta.
\sperc{64} of Wikimedia responses are less specific than a /24.
Google responds in \sperc{48} of cases with a prefix length less specific than 24, and Meta does so in \sperc{43} of cases.
It follows from these results that Wikipedia needs the fewest, and Meta has the largest number of queries.

While Meta does not use any scopes more specific than /24 prefixes, both, Google and Wikimedia do so.
RFC7871~\cite{rfc7871} does state that recursive resolvers should use a maximum prefix length of 24 to preserve user privacy.
These more specific responses do not affect resolvers adhering to RFC suggestions (\eg Google Public DNS and the OpenDNS) as they cache responses only for the full /24.
Google's own resolver also refuses to forward recursive queries with \ecs scopes larger than 24, responding with an extended DNS error~\cite{rfc8914} pointing to their guidelines~\cite{googleecsguidelines}.

\subsection{IPv6 Results}
\label{ssec:resp-aware-v6}

As described in \Cref{sec:approach}, we cannot simply perform a full address space scan in IPv6 similar to IPv4.
Instead, we use a prefix-seeded randomization approach with a scope prefix length of 48.
Similar to IPv4, we use the prefixes announced in BGP as seed list.
Additionally, we compare the BGP seed list to lists from \ac{irr} and /56 prefixes from the IPv6 Hitlist~\cite{gasser2018clusters, zirngibl2022rustyclusters}.
To evaluate the usefulness of such an approach, we compare it to scanning the different static lists of prefixes at the same time: BGP prefixes obtained from a local border router on April 26, 2024 (\sk{189}), prefixes in \texttt{route6} and \texttt{inet6num} objects from all RIRs' \ac{irr} databases on April 26, 2024 (\sm{1.3}), and all /56 prefixes from the responsive IPv6 Hitlist address set from April 1, 2024 (\sm{6.0}).
We conduct our scans with a single list of prefixes combined from all sources but evaluate the results for each prefix list source individually in the following.
The complete list of prefixes contains \sm{7.0} distinct IPv6 prefixes used as \ecs{} subnet parameters in our scan.

\begin{table}
    \setlength{\tabcolsep}{2.8pt}
    \caption{Comparison of the response-aware IPv6 approach with the prefix list (\sm{7.0}) comprised of BGP (\sk{189}), IRR (\sm{1.3}), and the IPv6 Hitlist (\sm{6.0}) prefixes. We shorten prefixes to /56 where necessary. \texttt{www.primevideo.com} and \texttt{tiktok.com} do not return \texttt{AAAA} records.}%
    \label{tbl:approach_comp}
    \footnotesize
    \centering
    \begin{tabular}{lrrrrrrrrrr}
    \toprule
     & \multicolumn{2}{c}{Resp.-aware} & \multicolumn{2}{c}{BGP} & \multicolumn{2}{c}{IRR} & \multicolumn{2}{c}{IPv6 Hitlist} & \multicolumn{2}{c}{Comb. Lists} \\
     \cmidrule(lr){2-3} \cmidrule(lr){4-5} \cmidrule(lr){6-7} \cmidrule(lr){8-9} \cmidrule(lr){10-11}
     & \#Qs & \#Addrs  & \#Qs & \#Addrs  & \#Qs & \#Addrs  & \#Qs & \#Addrs & \#Qs & \#Addrs  \\
    \midrule

    en.wikipedia.org  & \sm{1.3} &     \num{7}  & \sk{189} &     \num{6} & \sm{1.3} &     \num{6} & \sm{6.0} &     \num{6} & \sm{7.0} &     \num{6}  \\
    \midrule
    m.facebook.com    & \sm{1.6} &   \num{137}  & \dittostraight &   \num{137} & \dittostraight &   \num{137} & \dittostraight &   \num{137} & \dittostraight &   \num{137}  \\
    web.whatsapp.com  & \sm{1.6} &   \num{137}  & \dittostraight &   \num{137} & \dittostraight &   \num{137} & \dittostraight &   \num{137} & \dittostraight &   \num{137}  \\
    www.instagram.com & \sm{1.6} &   \num{137}  & \dittostraight &   \num{137} & \dittostraight &   \num{137} & \dittostraight &   \num{137} & \dittostraight &   \num{137}  \\
    \midrule
    www.google.com    & \sm{1.3} &  \num{1823}  & \dittostraight &  \num{1803} & \dittostraight &  \num{1739} & \dittostraight &  \num{2057} & \dittostraight &  \num{2058}  \\
    www.youtube.com   & \sm{1.3} &  \num{1587}  & \dittostraight &  \num{1556} & \dittostraight &  \num{1505} & \dittostraight &  \num{1690} & \dittostraight &  \num{1691}  \\
    \midrule
    www.amazon.com    & \sm{1.8} & \num{23266}  & \dittostraight & \num{23610} & \dittostraight & \num{25917} & \dittostraight & \num{28756} & \dittostraight & \num{28973}  \\
    \bottomrule
    \end{tabular}
\end{table}

\noindent\textbf{Scan Approach Comparison:}
\Cref{tbl:approach_comp} compares the response-aware IPv6 scanning approach with the static prefix list approach.
Due to ethical considerations, we apply query thresholds (see \Cref{sec:appendixquerylimits}) to limit the number of queries in the worst case to the same number as with the full IPv4 scan.
In contrast to IPv4, our scanner executes more queries for Meta domains than Google domains.
While the number of queries for Wikipedia also increases, our limits prevent the scan for Amazon from having as large a number of queries as for IPv4.
With the help of these limits, we keep the number of queries well below the ones performed by the prefix list scan.

Meta returns \num{137} addresses, and both approaches uncover the complete set.
We find \sk{1.8} and \sk{1.6} addresses for Google and YouTube, respectively, with fewer requests compared to Meta domains with our response-aware approach.
However, for Google domains, the approach with the static prefix list reveals \sperc{12.9} more IPv6 addresses but also requires more than five times as many queries.

As already seen in the IPv4 analysis, Google and YouTube also use multiple addresses inside a single response for \texttt{AAAA} queries.
Only Amazon changes its behavior from a 1:1 mapping in IPv4 to a 1:n mapping in IPv6.
We compare the efficiency of the two main scan approaches by evaluating the number of queries needed to uncover a new \ac{rrset} (\ie \acp{rrset} per query).
The response-aware approach needs only \sperc{25} of queries per \ac{rrset} compared to the prefix list scan for Google and \sperc{33} for YouTube, indicating a query reduction of $3\times$ and $2\times$, respectively.
In absolute terms, the response-aware approach uncovers more \acp{rrset} than the BGP and the IRR list.
Only the IPv6 Hitlist prefixes uncover more \acp{rrset}, at the cost of using \sperc{262} (\sm{4.7}) more queries.

In contrast to the \ac{rrset} results, which are an indicator of the usability of our scan for our load balancing goal, the infrastructure uncovering goal focuses on the number of distinct addresses obtained.
Therefore, we first analyze the different prefix sources in \Cref{tbl:approach_comp} and find that the IPv6 Hitlist alone finds all but one address, but it is also the largest source of prefixes (\sm{6} prefixes).
The IRR prefix list has more \ecs target prefixes than the BGP list but reveals fewer server addresses for both Google domains.
The BGP prefix list (\sk{189}) still identifies \sperc{97} of addresses collected by the response-aware approach and \sperc{88} of all addresses with only \sperc{3} of queries.
Therefore, we analyze the addresses unique to the IPv6 Hitlist prefix queries to understand the impact and whether we can create a limited set of \ecs prefixes to uncover the domain infrastructure efficiently.
Moreover, such an improved prefix set can be used as a seed set for the response-aware approach.

\noindent\textbf{Google:} \sperc{100} of the \texttt{www.google.com} server addresses unique to the IPv6 Hitlist prefixes are caused by prefixes announced by Cloudflare.
With \texttt{www.youtube.com}, these Cloudflare prefixes cover \sperc{72} of the unique addresses in responses.
The other \sperc{28} of prefixes are inside Google's own AS396982.
According to WHOIS information, the addresses in this \ac{as} are used by Google Cloud customers.
A closer look at the Cloudflare prefixes reveals that the IPv6 Hitlist contains \sk{30} /56 prefixes inside \texttt{2a09:bac2::/32} and \texttt{2a09:bac3::/32}.
Further investigation reveals that the hitlist addresses inside these prefixes are iCloud Private Relay~\cite{privaterelay} egress nodes operated by Cloudflare.
iCloud Private Relay is a two-hop relay service offered by Apple to protect the user's privacy in the network.
The egress nodes are operated by third parties such as Cloudflare.
iCloud Private Relay uses these addresses for traffic on the path between the service's egress and the destination servers.
Apple publishes a list of all egress ranges used by iCloud Private Relay as a geolocation feed~\cite{egress-nodes}.
Therefore, Google seems to use Apple's geolocation feed to optimize traffic for iCloud Private Relay users by sending dedicated responses to them.

Consequently, we employ a new list incorporating other IP geolocation feeds (geofeeds)~\cite{rfc8805,livadariu2024geofeeds}.
We gather feeds using the \emph{geofeed-finder}~\cite{geofeedfinder}, which leverages RFC9092~\cite{rfc9092} to identify geofeeds.
The initial test with the combined set (\sk{189} BGP, \sk{26} iCloud Private Relay, \sk{48} geofeed prefixes trimmed to /56) yields the same set of Google server addresses as the combined prefix list scan.
Therefore, combining BGP, iCloud Private Relay, and geofeed prefixes (\sk{245}) proves highly effective for \ecs scans.
Geofeeds, a relatively recent concept, could become a more valuable resource for future \ecs scans.

\noindent\textbf{Amazon:} The prefix list scan reveals \sperc{24.5} more addresses than our response-aware approach for Amazon.
\sk{28.2} of the returned IPv6 addresses are within the fully responsive prefix (FRP) dataset provided by the IPv6 Hitlist service~\cite{ipv6hitlist}.
The remaining addresses are in seven /48 prefixes and are not part of the IPv6 Hitlist.
AWS's encompassing /32 prefix contains \sm{1.2} FRPs of varying length.
We conduct an FRP detection for these additional prefixes based on the approach by \citeauthor{gasser2018clusters}~\cite{gasser2018clusters}.
Five out of the seven /48 are identified as such.
We could likely find more specific prefixes for the remaining two, which would also be fully responsive based on the existing data.

\noindent\textbf{Scope Prefix Lengths:}
While Amazon uses a static scope prefix length of 48, \sperc{51} of Meta's responses use a maximum scope prefix length of 32 bits (see Appendix \Cref{fig:scopesv6}).
Google and Wikipedia have \sperc{62} and \sperc{92} of responses scoped at 32 bits length or more specific.
This result shows that Meta provides much more fine-grained responses in IPv6 than the other two and, therefore, needs the largest number of queries from the three (\cf \Cref{tbl:approach_comp}).
Our scanner can be configured to perform fine-grained queries as necessary to the respective use case.
In our evaluation, we want to provide an overview of different deployments and, therefore, decide not to increase the prefix limits.
Future work can adapt these limits to obtain even more relevant data.
Similar to the observations for IPv4, we find that Google and Wikipedia return scope prefix lengths larger than 56, while Meta does not.
Google even returns scope prefix lengths of 128, and---again similar to IPv4---its own resolver refuses query scopes that are more specific than /56.

\subsection{Key Takeaways}
Our scanning approach reduces the number of queries needed for a full IPv4 address space scan by \sperc{86} to \sperc{97}, depending on the domain.
Reducing queries lessens the burden on the nameservers and enables timelier follow-up actions (\eg application layer scans).
In IPv6, we find an efficient prefix list consisting of BGP prefixes and prefixes extracted from geofeeds to uncover all observed addresses.
The analysis of the response scope prefix lengths shows that we do not have a snapshot of all relevant client-to-vantage point pairs, even if we obtain all observed distinct addresses.
Our scanner can use the prefix list as a seed to obtain in-detail insights into the nameserver's mappings.
Moreover, its efficiency in analyzing load balancing behavior is better than that of the prefix list.

\section{Evaluation of the ECS Landscape}
\label{sec:ecslandscape}
To confirm previous work and expand the scope of their analysis, we analyze the usage of \ecs in the wild.
This analysis helps us to understand the prevalence of nameservers and domains supporting \ecs and what researchers can expect when using our tool.
We perform and evaluate two types of scans:
First, we send four queries per IP protocol version to all authoritative nameservers of domains in the Cloudflare Radar~\cite{radar}, CrUX~\cite{crux}, Majestic Million~\cite{majestic}, and Umbrella~\cite{umbrella} domain lists (\Cref{ssec:exploration}).
Second, we conduct a more in-depth scan with client prefixes from all countries and a diverse set of \acp{as} to get a more detailed view of ECS-enabled domains.

\subsection{ECS Usage on the Internet}%
\label{ssec:exploration}

\begin{table*}
    \caption{Top 10 nameserver \acp{as} for \texttt{A}/\texttt{AAAA} queries. For CNAMEs, we follow the chain and query the resulting domain. Since domains can share the same canonical name, the set of queried domains, \textit{cDomains} is smaller. Shares are relative to Total. Domains can have nameservers in multiple \acp{as} (shares do not add up to \sperc{100})}. %
    \label{tbl:top10ecsnsas}
    \footnotesize
    \centering
    \begin{tabular}{llrrrrrrrr}
    \toprule
    && \multicolumn{4}{c}{IPv4 ECS Family} & \multicolumn{4}{c}{IPv6 ECS Family} \\
    \cmidrule(lr){3-6} \cmidrule(lr){7-10}
    Name & ASN & \#NS & cDomain & Domain & \ecs used & \#NS & cDomain & Domain & \ecs used \\
    \midrule
    Cloudflare & 13335 & 8024 & \sperc{73.2} & \sperc{76.9} & \sperc{2.2} & \num{7400} & \sperc{86.4} & \sperc{89.9} & \sperc{3.9} \\
    Amazon & 16509 & 4188 & \sperc{10.5} & \sperc{9.7} & \sperc{64.9} & \num{4092} & \sperc{4.4} & \sperc{3.9} & \sperc{53.2} \\
    China Mobile & 9808 & 109 & \sperc{3.3} & \sperc{3.5} & \sperc{0.4} & \num{67} & \sperc{0.1} & \sperc{0.0} & \sperc{0.6} \\
    Google & 15169 & 48 & \sperc{5.0} & \sperc{3.5} & \sperc{9.2} & \num{32} & \sperc{6.7} & \sperc{4.7} & \sperc{21.6} \\
    Kurun.com & 8796 & 28 & \sperc{3.2} & \sperc{3.4} & \sperc{0.0} & -- & -- & -- & -- \\
    Chinanet & 134763 & 25 & \sperc{3.2} & \sperc{3.4} & \sperc{0.0} & -- & -- & -- & -- \\
    Alibaba & 37963 & 238 & \sperc{2.3} & \sperc{2.3} & \sperc{2.5} & \num{158} & \sperc{0.2} & \sperc{0.2} & \sperc{2.0} \\
    Alibaba & 45102 & 86 & \sperc{1.3} & \sperc{1.1} & \sperc{1.9} & \num{42} & \sperc{0.2} & \sperc{0.1} & \sperc{0.7} \\
    Alibaba & 134963 & 24 & \sperc{0.9} & \sperc{0.8} & \sperc{0.5} & \num{18} & \sperc{0.1} & \sperc{0.1} & \sperc{0.1} \\
    Incapsula & 19551 & 124 & \sperc{0.8} & \sperc{0.7} & \sperc{5.1} & -- & -- & -- & -- \\
    \midrule
    Total & & \sk{15.8} & \sm{1.1} & \sm{1.2} & \sk{184} & \sk{13.0} & \sk{794} & \sk{853} & \sk{64} \\
    \bottomrule
    \end{tabular}
\end{table*}

During this first scan, we send DNS queries for each domain to their authoritative nameservers.
We send individual queries with \ecs{} information for four IPv4 prefixes and four IPv6 prefixes.
We select a diverse set of prefixes based on location and topology (see \Cref{sec:appendixquerylimits}).
{Response-aware} address space scans, as evaluated in \Cref{sec:response-aware}, show that this combination provides the best coverage of locations and corresponding distinct responses.

\noindent\textbf{Scanning Setup:} \ecs only provides value to domains hosted on distributed deployments, so we focus on domains in top lists.
These popular domains usually require some sort of load balancing to handle the received load.
With this measurement, we want to understand how many domains and nameservers enable and use \ecs{}.
The selected lists use different collection approaches (DNS-based for Cloudflare Radar and Umbrella; web statistics/user-based for Majestic and CrUX).
Therefore, they cover a large set of relevant domains.
The four domain lists have \sm{3} combined unique domain names with resolvable authoritative nameservers.

We limit our scan to \texttt{A} queries with IPv4 client subnets and \texttt{AAAA} queries with IPv6 clients.
If the domain name resolves to a \texttt{CNAME}, we follow the chain and perform our \ecs queries towards the authoritative nameserver of the resulting domain name.
Multiple domains of our input set can resolve to the same canonical name, \eg{} domains hosted by Cloudfront.
For those domains, load balancing based on \ecs{} is done for the A or AAAA resolution of the canonical name.
Thus, we test the behavior for those domains, reducing our set of targeted domains to \sm{2.7} domain names.
In the following evaluation, we refer to the original input set based on the top lists as \textit{Domains} while referring to the scanned set of domains with resolved canonical names as \textit{cDomains}.

\noindent\textbf{ECS Support in the Internet:}
The authoritative nameservers of the analyzed domain names are distributed across \sk{196} IPv4 and \sk{39} IPv6 addresses.
These nameserver IP addresses are located in \sk{21} \acp{as} (\sk{20.8} IPv4 and \sk{5.1} IPv6).
These numbers show that our target set is diversified over many \acp{as}.
\sm{1.1} \textit{cDomains} (\sperc{41}, \cf \Cref{tbl:top10ecsnsas}) have \ecs enabled on their authoritative nameserver. %
These \textit{cDomains} are used by \sm{1.2} domains from our initial input set.
This result is similar to what \citeauthor{kountouras2021understanding}~\cite{kountouras2021understanding} found in 2019.
While we receive relatively fewer successful responses to our IPv6 \ecs queries, our scan still reveals \sm{0.8} \textit{cDomains} (\sperc{30}) supporting \ecs{}.
In total, only \sk{8.9} IPv4 (\num{449} \acp{as}) and \sk{6.9} IPv6 (\num{140} \acp{as}) nameserver addresses (\sk{15.8} in total) are authoritative for these ECS-enabled \textit{cDomains}.
These are only \sperc{5} and \sperc{18} of queried IPv4 and IPv6 nameservers, respectively.
Only \sperc{6} of domains (\sperc{2} for \ecs with IPv6) respond with at least two different \acp{rrset} (see \Cref{tbl:top10ecsnsas}).
This is \sperc{15} and \sperc{11} of \ecs-enabled domains for IPv4 and IPv6, respectively.
We use this as an indicator of whether \ecs is actually used.
Deploying \ecs{} has preconditions (\eg{} a distributed setup) and requires substantial effort. %

\noindent\textbf{Nameserver Operators:}
\Cref{tbl:top10ecsnsas} lists the 10 \acp{as} containing the nameservers authoritative for the largest number of domains that provide \ecs support.
Cloudflare serves \sperc{76.9} of all domains supporting \ecs over IPv4 and \sperc{89.9} over IPv6.
However, only a small share of these domains result in two different answers for the four queries per IP version.
To better understand how nameservers use \ecs{}, we perform a detailed analysis of more specific scans in \Cref{ssec:moreprefixeval}.
Kurun.com, Chinanet, and Incapsula are the only \acp{as} in the top ten that do not support \ecs with AAAA queries.

By a substantial margin, AWS has the largest share (\sperc{65}) of domains using \ecs with more than one answer.
AWS provides a simple configuration mechanism that allows its users to configure the exact answers per client subnet~\cite{route53_ecs}.
This simple and fully customizable configuration, together with the primary cloud products by AWS, might be an important factor for the large number of \ecs-using domains with AWS.
The \ac{as} responsible for the second-largest share of such domains is Google, with only \sperc{9} of domains, followed by Incapsula with \sperc{5}.

\sperc{43} of IPv4 nameservers and \sperc{53} of IPv6 nameservers enable and use \ecs only for select domains.
\sperc{55} of the responsible nameservers are within the prefixes \texttt{205.251.192.0/21} and \texttt{2600:9000:5300::/45}.
Both prefixes are owned by AWS and used for Route53~\cite{aws_ranges}.
They are identified as being fully responsive~\cite{zirngibl2022rustyclusters, sattler2023hrps}, \ie all usable addresses inside these prefixes are responsive.
The second largest \ac{as} is Cloudflare, covering an additional \sperc{27}.
It is reasonable for these providers to enable and use \ecs only for select domains as they charge for the service, and not every domain owner uses this feature.

\subsection{Scanning with More Client Subnets}%
\label{ssec:moreprefixeval}

To better understand how nameservers make use of \ecs{} and to extend our previous scan to identify domains and nameservers generally supporting \ecs{}, we perform a scan with more prefixes.
Due to the large number of domains, we cannot perform a full address space scan for each.
If available, we select three prefixes per ISO 3166 country code~\cite{iso3166} to obtain a geographically and topologically diverse set of prefixes.
We use data provided by IPinfo~\cite{ipinfo} to geolocate all routed IPv4 /24s by their network address.
For IPv6 we used /48s from our seed lists (BGP, IRR, and IPv6 Hitlist~\cite{gasser2018clusters,zirngibl2022rustyclusters}).
Our algorithm also ensures never to select prefixes originated by the same \ac{as}.
These two decisions provide us with a geographically and topologically diverse sample set to efficiently obtain an overview of how nameservers use \ecs{} without putting too much load on them.
This selection algorithm results in 698 IPv4 and 609 IPv6 prefixes.

\Cref{tbl:top10ecsnsas} shows that only \sk{184} out of \sm{1.2} domains (for IPv6 \sk{64} out of \sk{853}), thus less than \sperc{10}, use the \ecs option (\ie we observe more than a single \ac{rrset}).
In contrast to the scans in \Cref{ssec:exploration}, in this scan, we use 698 prefixes to determine if nameservers actually return more than one \ac{rrset}.
We perform this additional scan for all domains that appeared to use \ecs during our initial scan.
Additionally, we scanned a sample of \sk{40} \ecs enabled but not using domains.
We focus our evaluation on IPv4 results as IPv6 results are similar.

\begin{figure}
    \includegraphics{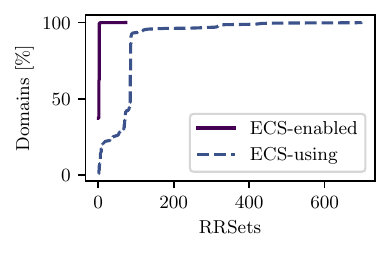}
    \centering
    \caption{ECDF for the number of observed \acp{rrset} in the country prefix list scan.}%
    \label{fig:moreprefixanswers}
\end{figure}

We argue that if we only observe a single \ac{rrset} for this larger \ecs prefix set for a given domain, the domain is not using the \ecs option in a way to the benefit the user.
Moreover, the resolver must also provide needless caching space for these domains.
Therefore, the user ends up losing its privacy to the authoritative nameserver without any gains while the resolver's caches suffer.

\noindent\textbf{ECS-enabled Analysis:}
In \Cref{fig:moreprefixanswers}, we show the cumulative distribution of domains and their corresponding number of observed \acp{rrset}.
For \sperc{37} of the sampled \ecs-enabled domains, we still observe only a single set of returned records.
\sperc{66} of these domains have a Cloudflare authoritative nameserver and \sperc{10} a Google nameserver.
This seems counterintuitive at first glance, as \ecs load balancing is a paid service according to the documentation~\cite{cloudflareecs}, but the domains seem to not make use of it.
We use a free tier Cloudflare account with a registered domain name and find that Cloudflare's authoritative nameservers return scoped answers even though we did not enable it.
Since our domain cannot actually make \emph{use} of this feature without paying for it, our domain does not benefit from this behavior.
When considering this Cloudflare specialty, we find that \sperc{99.7} of sampled ECS-enabled domains are not \emph{using} \ecs{}.

\noindent\textbf{ECS-using Analysis:}
More than \sperc{99} of domains that appeared to use \ecs in our initial scan also did so in this follow-up scan (see \Cref{fig:moreprefixanswers}).
Cloudflare's behavior of few distinct responses is present but only marginally visible (less than \sperc{1}).
\sperc{20} of domains have ten or fewer distinct record sets.
We observe two significant jumps: The first and smaller one contains \sperc{12} of domains with 67 to 73 \acp{rrset}, and the second one between 81 and 88 sets covers \sperc{50} of all scanned domains.
AWS authoritative nameservers cause both.
The records also point to addresses inside the AWS address space.
This leads us to the conclusion that these are two different default AWS configurations to perform global DNS-based load balancing using the Amazon CloudFront CDN.

Besides the set of ECS-enabled domains with only a single response, Cloudflare is also responsible for \sperc{99} of domains with two or three distinct \acp{rrset}.
Interestingly, most domains with a single \ac{rrset} do not resolve to Cloudflare IP addresses, while Cloudflare announces more than \sperc{99} of the addresses from domains with multiple \acp{rrset}.
This could indicate that Cloudflare actively uses \ecs{} for domains hosted within their network.
However, according to data by IPinfo~\cite{ipinfoanycast}, except for one, all IP addresses returned by Cloudflare are inside IP anycast prefixes.
Furthermore, our scan reveals only three \acp{rrset}, while Cloudflare has more than 300 global \acp{pop}~\cite{cloudflare_network}.
Therefore, we assume that domains with two and three responses are also not actually based on \ecs{} information.
Instead, we assume that Cloudflare is balancing our requests using a strategy that does not rely on \ecs{}.
The observed \ecs{} behavior would then be similar to the ECS-enabled domains.
However, this behavior can fill up the cache of ECS-using recursive resolvers, and it allows Cloudflare to collect data on all request-originating clients' subnets of such a resolver.

\subsection{Key Takeaways}
\reviewfix{3}
Our evaluation finds that \sperc{40} of the analyzed top-list domains have \ecs enabled for IPv4 subnets on their authoritative nameservers, but these domains are served by only \sperc{5} of queried nameserver IP addresses.
The \ecs deployment is highly skewed towards large DNS providers.
Scanning with four prefixes revealed that only \sperc{15} of these domains actually use \ecs{}.
A scan with 698 ECS prefixes validated these results and identified Cloudflare as the major player responsible for the delta between domains with \ecs enabled and those using it.
Cloudflare enables \ecs for domains even if they do not pay for the service, but it will not tailor the response records for these.
This result shows that a single query is not representative of determining if \ecs is actually used.
According to our results, four prefixes are sufficient to precisely categorize ECS-using domains.

\section{Validating ECS Properties of Nameservers}
\label{sec:nsproperties}

Our results show that \ecs is used by a considerable number (\sk{184}) of domains.
Our scanning approach helps collect addresses and understand the domains load balancing behavior.
As part of this section, we examine important properties of the domains authoritative nameservers.
These findings should be considered when performing \ecs scans and analyzing its results.

\subsection{Public Resolvers vs Authoritative Nameservers}
\label{ssec:pubresolver}

Related work~\cite{warrior2017drongo,aldalky2019ecs} mentioned that some authoritative nameservers provide \ecs responses only to a restricted set of resolvers.
We re-evaluate this statement to validate if we can directly query authoritative nameservers instead of resolvers.
This would have the ethical positive side effect of not unnecessarily filling up the public resolvers' caches.
As we aim to find the most ethical scanning approach, we evaluate the benefits and drawbacks of using public resolvers.
Therefore, we issue the same queries to Google's public resolver and authoritative nameservers.
We chose Google's resolver as it is one of the most well-known ECS-supporting public resolvers, and it was listed as one of the few with access to restricted deployments~\cite{warrior2017drongo}.
Thus, if a service deploys a restricted \ecs setup to its authoritative nameservers, it will most likely enable it for Google's resolver.

We limit this scan to a single IPv4 \ecs prefix and to domain names in Cloudflare Radar and CrUX (\sm{2.1} domains in total) to reduce the load on the resolver.
A single query is sufficient to observe if \ecs responses of certain domains are restricted to an allowed set of public resolvers.
\sk{824} domains (\sperc{40}) return a scoped answer either on the authoritative nameserver scan or the recusive resolver scan or both.
\sk{2.2} (\sperc{0.3}) of these ECS-enabled domains provide \ecs information only to Google's public resolver and not to our direct queries to their authoritative nameservers.
The Soprado GmbH (AS20546) is the most important provider, with \num{230} domains (\sperc{10}) returning \ecs scopes only to queries from Google Public DNS.
On the other hand, for \sk{5.2} (\sperc{0.6}) domains, our scanner reports scoped answers while the Google resolver does not.
Cloudflare's authoritative nameservers are responsible for \sk{3.9} of these domains with missing \ecs{} information in the Google resolver's responses.
The information is either not added by the nameservers or removed by the resolver.
Google's guidelines list several rules a nameserver has to follow for Google to send \ecs queries (\eg overlapping \ecs responses, as analyzed in \Cref{ssec:overlapping}).
This evaluation shows that the prevalence of restricted \ecs deployments is no longer a widespread phenomenon.

However, we wish to emphasize a scenario where recursive resolvers prove more advantageous as targets: \texttt{CNAME} chain resolution.
Unlike our scanner, which issues specific queries but does not perform resolution tasks required for \texttt{CNAME} records\footnote{
In addition to our \ecs scanner \emph{ECSplorer}~\cite{ecsplorer}, we release a tool to perform \texttt{CNAME} resolution and obtain the authoritative nameservers of the domain to query.
The tool produces an input list for the \ecs scanner to use.
The drawback compared to a recursive resolver is that this resolution is not done as part of each \ecs query, and we observe rare cases where nameservers change behavior during scanning periods.}, a recursive resolver can resolve \texttt{CNAME} chains.
While this ensures receipt of the desired record type for the domain name, it also relinquishes control over the resolution path, leaving us unaware of the authoritative nameservers employed.
For \texttt{CNAME} resolution needs, one could opt for a local recursive resolver on the scan machine as an alternative to public resolvers.

In summary, the prevalence of restricted \ecs deployments is negligible (\sperc{0.3}).
Therefore, the necessity of using public resolvers is generally not given anymore.
Directly querying the authoritative nameserver reduces the impact on and by resolver caches.
Nevertheless, the use of recursive resolvers should be considered when \texttt{CNAME} resolutions are relevant.

\begin{figure*}
    \begin{subfigure}{.345\textwidth}
        \centering
        \includegraphics{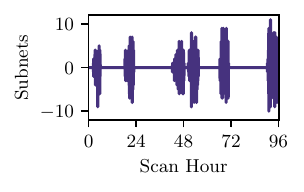}
        \caption{www.google.com}%
        \label{fig:scopechangesgoogle}
    \end{subfigure}
    \begin{subfigure}{.325\textwidth}
        \centering
        \includegraphics{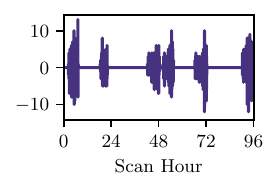}
        \caption{www.youtube.com}%
        \label{fig:scopechangesyoutube}
    \end{subfigure}
    \begin{subfigure}{.315\textwidth}
        \centering
        \includegraphics{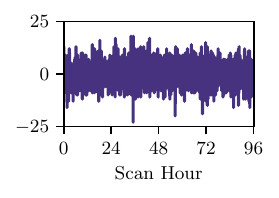}
        \caption{www.instagram.com (Meta)}%
        \label{fig:scopechangesmeta}
    \end{subfigure}
    \caption{Scope prefix changes observed in 96\ h. Positive values are changes to more specific responses and negative ones to less specific. The scan started on Friday, May 10, 2024, at 15:45 UTC, at our vantage point in Germany. See \Cref{fig:scopechanges6} for IPv6 plots.}%
    \label{fig:scopechanges}
\end{figure*}

\subsection{Impact of Scanning Location}
To validate whether a single vantage point is representative to uncover a service's infrastructure (the first goal of our approach), we perform parallel scans from four additional vantage points.
While our main scanning location is in an academic network in Western Europe (Germany), we added another academic one in Europe (France) and three vantage points from a cloud provider in Frankfurt - Germany, Singapore, and Newark, New Jersey - United States.
We start the scans at the same time, ensuring the same query order.

\noindent\textbf{Top List Scan Impact:}
Due to the larger number of concurrent queries, we limit the scan to Cloudflare Radar domains and four IPv4 \ecs prefixes.
\sk{380} Cloudflare Radar domains have \ecs{} enabled.
For \sperc{31} of these, we observe at least two different \acp{rrset} at different vantage points.
\sperc{98} of these domains with multiple \acp{rrset} are served by Cloudflare (AS13335), AWS Route53 (AS16509), or Meta (AS32934) authoritative nameservers.
The responses for these names contain different nameserver identifier (NSID)~\cite{rfc5001} option values at different vantage points%
\footnote{For Meta, we decoded the Base64-encoded value to extract the cluster value. This is necessary as Meta nameservers encode a timestamp and the \ecs value inside the NSID option value.},
indicating that we hit different instances of an IP anycast deployment.
Nevertheless, this affects only \sperc{7} of all domains using an AWS Route53 nameserver and \sperc{34} of domains served by Cloudflare nameservers.
Cloudflare exhibits a unique behavior: while it provides scoped answers, the response varies based on the requester's IP address and not the client subnet.
This finding is analogous to \Cref{ssec:moreprefixeval}, where we scan from a single vantage point, and most domains managed by Cloudflare do not effectively use \ecs{}, always responding with the same \ac{rrset}.

To evaluate this hypothesis, we use CAIDA's Archipelago Globally Distributed Active Measurement Platform (Ark)~\cite{caida-ark}, which offers hundreds of vantage points around the globe and has the ability to send \ecs queries~\cite{scamper}.
We reimplemented our scanning approach~\cite{ecsplorer-ark} in Python to make use of Ark.
The validation scan uses \num{130} Ark vantage points to issue a predefined DNS \ecs query from all vantage points towards the same authoritative nameserver IP address.
Across a set of 30 different subnets in the \ecs extension, Cloudflare selects the same IP address for a specific vantage point disregarding the actual \ecs information, while claiming the response is only valid for the specific \ecs subnet.
This behavior is only visible for domains which are actually load balanced via DNS.
These results confirm our hypothesis on Cloudflare's behavior of ignoring \ecs information.

\noindent\textbf{Address Space Scan Impact:}
We use a distributed full IPv4 address space scan (same methodology as in \Cref{sec:response-aware}) to evaluate the prevalence of mismatches within such a scan.
The answers from Google's nameservers match for \sperc{97} of all queries at all locations.
For Meta domains, \sperc{81} of queries, and for Amazon, only \sperc{68} of queries return an equivalent response.
Disregarding the concrete responses to queries, we uncover the same aggregated set of server IP addresses at all locations for Google and Meta.
Amazon provides some IP addresses only to certain locations.
Nonetheless, we find that the reverse DNS names of the locations point to the same set of \acp{pop} (see also \Cref{ssec:geolocation}).
As Amazon is known to have a distinctive use of its address space~\cite{ipv6hitlist} to balance the load, we attribute our observation to this behavior.

\noindent\textbf{Infrastructure Coverage Validation:}
We issue RIPE Atlas A record DNS measurements using all available probes to obtain results from up to \sk{11.6} vantage points.
We schedule the measurements to run simultaneously as our response-aware scan.
Before evaluating the results, we remove any bogus answers.
The measurements from probes located in China return Meta-owned addresses.
The behavior results from the Great Firewall of China (GFW) injecting forged answers for queries carrying censored domains~\cite{steger2023targetacquired,hoang2021great}.
Therefore, we do not count these as real answers by the authoritative nameserver and disregard them in our comparison.

Amazon answers have already shown a distinctive load balancing behavior depending on the location.
We validate Wikipedia answers using their published dataset~\cite{wikipediadcs}.
Thus, we limit this validation to Google and Meta.
For Meta, we find 135 IPv4 and IPv6 addresses, all of which are seen in our response-aware scan.
The measurements for Google and YouTube collect \sk{1.2} addresses for each domain for both IPv4 and IPv6.
Therefore, we covered \sperc{33} fewer addresses than our response-aware scan while not contributing any previously unknown addresses.

In summary, our results show that the two most important \acp{as} do not provide stable responses when evaluating a distributed scan.
However, this only affects \sperc{31} of ECS-using domain names.
Our full address space analysis also shows differences in the responses depending on the scanning location.
Nevertheless, from an infrastructure coverage view it finds as many addresses as actual distributed scans do.

\subsection{Stability of Responses}
\label{ssec:stability}

As previously established, the purpose of \ecs is to balance the load based on client prefixes.
Load balancing is typically a dynamic process involving several aspects, making it a complex system to understand.
To validate how often responses change, we evaluate the load balancing update behavior by zooming in on changes in the returned scope prefix length.
We use the country prefix list to better understand when and how the scopes change and run \ecs measurements every three minutes for 96 hours.
We limit this scan to Google, YouTube, and Instagram (Meta).
Meta behaves consistently the same way for all three observed domains, while Google and YouTube do not.

In \Cref{fig:scopechanges}, we visualize the scope changes over time.
While Meta changes at least some scope prefix length at every measurement, Google and YouTube have a very distinctive behavior.
Both exhibit changes only in specific time windows.
We find a daily pattern that is approximately four hours before the end of the measurement day and lasts until its end (\eg just before the 24th scan hour).
Two additional time frames start right after a new measurement day (after hour 0 and 48).
Neither the \ecs prefix's location nor topological properties seem to be relevant for a prefix to change the returned scope.
While we leave a detailed analysis of the load balancing behavior to future work, this evaluation shows that our data reveals new load balancing aspects.

\subsection{Overlapping Subnets}
\label{ssec:overlapping}

RFC7871~\cite{rfc7871} states that authoritative nameservers must not overlap prefixes.
Google's resolver rules~\cite{googleecsguidelines} also include this requirement and its resolvers will stop using \ecs{} with nameservers not adhering to it.
Overlapping responses are also counterintuitive as the less specific answer can overshadow the more specific one, resulting in inconsistent states between resolvers.

\Cref{tab:overlap_example} shows an example of the problem.
A resolver sends two queries with neighboring /24 prefixes (within the same /23).
The nameserver responds to the first query with a scope of 24.
However, it responds to the second one with a scope of 23, covering the first query's prefix scope.
This example leads to a caching issue at the resolver: Should it cache both answers or overwrite the first one?
As the RFC forbids such behavior, it does not contain a solution.
If the resolver first issues the second query, it will not issue the first query as long as the overlapping response is cached.

\begin{table}
    \centering
    \footnotesize
    \caption{Overlapping ECS: The response for query 2 can be cached for the /23 covering the first response.}
    \label{tab:overlap_example}
    \begin{tabular}{lllrrl}
        \toprule
        & Domain & ECS Address & Source & Scope  & Response \\
        \midrule
        1& \texttt{example.com} & \texttt{10.0.0.0} & 24 & 24  & \texttt{192.0.2.1} \\
        2& \texttt{example.com}  & \texttt{10.0.1.0} & 24 & 23 & \texttt{198.51.100.1} \\
        \bottomrule
    \end{tabular}
\end{table}

Overlapping responses are not allowed, yet we detected overlapping scopes in our initial scans for Google, YouTube, and all domains by Meta.
We can rule out any time-related effects in our results as our response-aware approach scans these prefixes sequentially within milliseconds.
AWS Route53 nameservers exclusively provide /24 or /48 scopes, thereby precluding any overlapping response scopes.
Conversely, we do not observe any overlapping prefix scopes for Wikipedia.

\begin{table}
    \setlength{\tabcolsep}{3.2pt}
    \caption{Overlapping response scopes in a full non-response-aware /24 subnets scan. SubASNs is the number of covering prefixes where at least one covered prefix is in a different AS. We only scan a single Meta domain as all behave the same.}%
    \label{tbl:sl24fullstats}
    \centering
    \footnotesize
    \begin{tabular}{lrrrrr}
    \toprule
     & Cov. Pfxs & SubASNs & Answ & /24s & BGP Pfxs \\
    \midrule
    www.instagram.com & 464 & 352 & 352 & 333 & 52 \\
    www.google.com & 1026 & 996 & 958 & 944 & 69 \\
    www.youtube.com & 5987 & 5760 & 5760 & 5533 & 779 \\
    \bottomrule
    \end{tabular}
\end{table}

In order to quantify the problem in a structural manner we perform an additional \ecs{} scan for all announced /24 prefixes.
Our response-aware approach only detects overlapping responses if the first query results in the more specific answer and only a later answer covers the previous one.
It skips the remaining \ecs{} targets if the first query returns the less specific answer.
In \Cref{tbl:sl24fullstats} we list the number of less specific answers covering more specific ones from other queries.
We perform this more detailed scan only for \texttt{www.instagram.com} on Meta's side as Meta responds consistently with the same scopes for all three observed domains.
In comparison, Google does behave differently for \texttt{www.google.com} and \texttt{www.youtube.com}.

We find that Meta has the smallest number of answers overlapping with more specifics.
YouTube has nearly \sk{6} prefixes with at least one more specific response that would be overlapped.
The responses provided are within different /24 subnets, and \sperc{13} of YouTube, \sperc{7} of Google, and \sperc{11} of Instagram prefixes respond with addresses in different BGP prefixes.
This result is especially outstanding as Google's guidelines lay out in detail that such behavior does not conform to RFC7871~\cite{rfc7871}.
Affected prefixes are even more interesting, considering that the two clients from our example could be in different \acp{as}.
We find that \sperc{76} of affected prefixes with Meta and more than \sperc{97} of affected prefixes with Google and YouTube contain more specific client prefixes assigned to a different AS than the covering prefix.
While the total number of affected prefixes is limited, it is important to consider these cases when performing full address space scan evaluations.

\subsection{Key Takeaways}
\reviewfix{3}
In this section, we used our scanning approach to analyze four aspects of nameserver \ecs behavior:
\first In contrast to previous work~\cite{warrior2017drongo,aldalky2019ecs}, we
find no significant indication of nameservers restricting their \ecs support to specific resolver addresses.
Therefore, researchers can directly query authoritative nameservers instead of using ECS-enabled resolvers, thereby reducing the footprint of \ecs scanning campaigns on third-party infrastructure.
\second Distributed measurements show that different IP anycast instances of nameservers provide distinct answers. While our scan significantly improves the visibility of single vantage point scans, it does not replace real distributed scans.
\third Higher frequency measurements reveal a scope update behavior by Google, which seems to follow daily patterns.
This finding shows potential for future research to uncover the load balancing behavior of ECS-enabled services.
\fourth We find nameserver operators---one of which is Google---that provide overlapping responses.
Google's public resolver has rules for \ecs responses by authoritative nameservers that prohibit such overlapping responses.
Therefore, Google's authoritative nameservers do not adhere to the rules defined by its own public resolver.

\section{Conclusion}
\label{sec:conclusion}

We present our novel response-aware \ecs measurement approach to perform full address space scans for IPv4.
It also includes a seeded response-aware variant to support IPv6 scanning.
Our approach can be used to uncover the infrastructure of domain names from a single vantage point and collect information on load balancing behavior.
It reduces the number of queries by up to \sperc{97} compared to previous work.
We find that \sperc{40} of top-listed domain names enable \ecs{}, but we only observe that \sperc{6} use it by providing tailored responses.
However, these domains contain important services managed by different operators.
Our presented approach allows us to evaluate their load balancing behavior in more detail and use results for future work. %
We find that Cloudflare, the largest nameserver provider in our dataset, responds with scoped answers but rarely provides subnet-tailored answers.
For Google, we find two of their own resolver's rules to be violated by their own authoritative nameservers.
We make our response-aware scanner (\url{https://github.com/tumi8/ecsplorer}~\cite{ecsplorer}) and data (\url{https://doi.org/10.14459/2025mp1779517}~\cite{datatum}) available to the public.

\begin{acks}

We thank the anonymous reviewers and our shepherd Siva Kesava Reddy Kakarla for their valuable feedback.
We would like to thank Matthew Luckie for
implementing the \ecs feature into the CAIDA Ark platform and providing support
for our supplementary scanning campaign (NSF OAC-2131987).
This work was partially funded by the German Federal Ministry of Education and
Research under project PRIMEnet (16KIS1370).
\end{acks}

\label{body}

\bibliographystyle{ACM-Reference-Format}
\bibliography{refs}

\appendix

\section{Ethics}
\label{sec:ethics}
All our scans are set up based on a set of ethical measures we follow strictly.
These are mainly based on informed consent~\cite{menloreport} and well known best practices~\cite{PA16}.
This study does not involve users, their information or sensitive data but focuses on publicly reachable and available services.
To not cause harm to any infrastructure, we apply measures described by \citeauthor{durumeric_zmap_2013} \cite{durumeric_zmap_2013}.
We limit the rate of our scans and use a blocklist based on requests to be excluded from our scans.
We are directly registered as abuse contact for our scan infrastructure and react quickly to all requests.
Furthermore, we host websites on all IP addresses used for scanning to inform about our research and provide contact information for further details or scan exclusion.
We did not receive any exclusion requests during this scanning campaign.

Our response-aware scans are designed to reduce the query load as much as possible.
Respecting the scope indicated by the nameserver allows to drastically reduce the query load compared to a scan of all /24 subnets as shown in \Cref{sec:response-aware}.
\section{Artifacts, Configurations, and Complementary IPv6 Results}
\label{sec:appendixquerylimits}

\noindent\textbf{Artifacts:} We will publish the source code, instructions, and used configurations for the \ecs scanner and the accompanying tool to follow CNAMEs to a public Git repository.
Additionally, we will push scanning result data to a public data archive.

\noindent\textbf{Configurations:} We use the following four IPv4 and IPv6 prefixes the \ac{ecs} exploration scan in \Cref{ssec:exploration}.
\begin{itemize}
    \item[IPv4:] \texttt{108.238.84.0/24}, \texttt{2.59.158.0/24}, \texttt{5.200.28.0/24}, \texttt{1.23.92.0/24}
    \item[IPv6:] \texttt{2600:1700::/48}, \texttt{2a00:1630::/48}, \texttt{2a01:6f0:100::/48}, \texttt{4000:2002::/48}
\end{itemize}

The prefix limits in \Cref{tab:limits} are applied to each trie node, which represents a specific prefix, with a specific length.
That means each /48 is allowed to have the configured amount of scans.
We use limits in our IPv6 address space scans, for our response-aware scan introduced in \Cref{sec:approach} and used in \Cref{sec:response-aware},
The first category applies to all prefixes and subprefixes in the given seed prefix set.
The prefix lengths where selected after an evaluation of BGP announced prefix lengths and their prevalence.
Additionally, we also allow to configure a combined (routed and not routed) limit per prefix.
In our case we set these values to the sum of routed and not routed limits.
While we also support limits for IPv4, full address space scans do not use any limits.

\begin{table}
    \caption{IPv6 scan limits for our response-aware scanner.}
    \label{tab:limits}
    \begin{tabular}{llllll|lll}
        \toprule
        & \multicolumn{5}{c}{Routed} & \multicolumn{3}{c}{Unrouted} \\
        \midrule
        Prefix Size & /48 & /40 & /32 & 29 & /16 & /32 & /16 & Total\\
        Queries & 1 & 4 & 64 & 512 & \num{32768} & 1 & 200 & \num{16384}\\
        \bottomrule
    \end{tabular}
\end{table}

\begin{figure*}
    \vspace{-1.5em}
    \begin{subfigure}{.347\textwidth}
        \centering
        \includegraphics{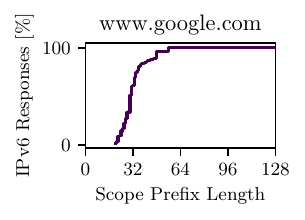}
    \end{subfigure}
    \begin{subfigure}{.315\textwidth}
        \centering
        \includegraphics{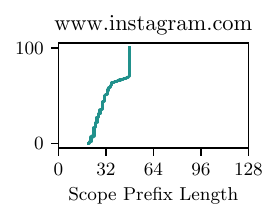}
    \end{subfigure}
    \begin{subfigure}{.315\textwidth}
        \centering
        \includegraphics{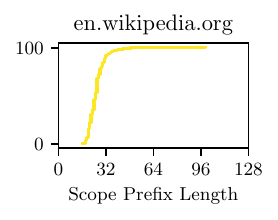}
    \end{subfigure}
    \caption{Scope prefix lengths observed at our IPv6 response-aware scans. AWS Route53 always uses /48. More information can be found in \Cref{ssec:resp-aware-v6}. For comparison, \Cref{fig:scopes} shows results for IPv4.}
    \label{fig:scopesv6}
\end{figure*}

\begin{figure*}
    \begin{subfigure}{.345\textwidth}
        \centering
        \includegraphics{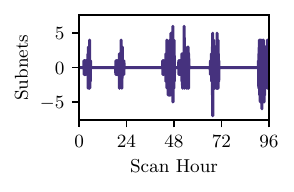}
        \caption{www.google.com}%
        \label{fig:scopechangesgoogle6}
    \end{subfigure}
    \begin{subfigure}{.325\textwidth}
        \centering
        \includegraphics{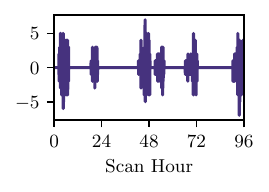}
        \caption{www.youtube.com}%
        \label{fig:scopechangesyoutube6}
    \end{subfigure}
    \begin{subfigure}{.315\textwidth}
        \centering
        \includegraphics{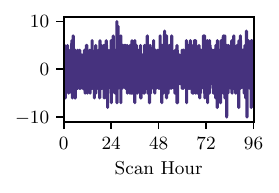}
        \caption{www.instagram.com (Meta)}%
        \label{fig:scopechangesmeta6}
    \end{subfigure}
    \caption{IPv6 Scope prefix changes observed in 96\ h. Positive values are changes to more specific responses and negative ones to less specific. The scan started on Friday, May 10, 2024, at 15:45 UTC, at our vantage point in Germany.}%
    \label{fig:scopechanges6}
\end{figure*}

\noindent\textbf{Complementary IPv6 Figures:} \Cref{fig:scopesv6} shows the scopes returned by authoritative nameservers for the respective domains in our response-aware scans as discussed in \Cref{ssec:resp-aware-v6}.
In comparison, results for our IPv4 scans can be found in \Cref{fig:scopes}.
\Cref{fig:scopechanges6} shows the changes of response granularity for IPv6 client addresses as discussed in \Cref{ssec:stability}.
IPv6 and IPv4 results (see \Cref{fig:scopechanges}) show changes in the same time frames for Google and Youtube.

\section{Analyses of ECS Research Use Cases}%
\label{sec:showcases}
We present use cases, including a basic analysis to show the usefulness of \ecs{} for research, besides the general use case to load balance based on client locations.
While these sections are not a central part to our contribution, they show potentially interesting future applications of our work.

Our response-aware scan can collect information on how clients are mapped to hyper-giant infrastructure.
Large providers put significant effort into this mapping, resulting in precise geolocation, sometimes even better than geolocation databases.
Furthermore, \ecs{} provides possibilities to conduct \textit{virtually distributed} scans.
Therefore, it helps to improve domain-dependent scans such as QUIC, allowing for more successful handshakes.
We further evaluate the usefulness of \ecs scanning for IPv6 hitlists and server deployment locations.

\subsection{Matching Server and Client Locations}
\label{ssec:geolocation}
\citeauthor{calder2013ecs}~\cite{calder2013ecs} presented the \emph{client-centric front-end geolocation (CCG)} approach.
CCG uses majority voting to determine the location of \emph{front-ends} (\ie the \ac{pop}'s IP addresses).
Their approach relies on the assumption that a nameserver will assign clients to destination addresses nearby.
In contrast to their approach, to identify the server locations, we suggest to use \ecs{} data to estimate client location based on the data collected by large providers.
We rely on location hints in the reverse DNS pointer record of the \ac{pop} IP addresses to estimate the location of \acp{pop} and the clients assigned to them.
We show the potential of our suggestion by comparing our estimates to IP to location databases, and showcase an example where \ecs{} provides a better estimate.

We assume that location hints in the rDNS name of servers can serve as a reliable indicator for mapping the server's location.
Even if the exact physical location might not be fully accurate, the rDNS location hint expresses a data point controlled and actively set by the provider.
We use our full address space scan results from \Cref{sec:response-aware} and manually determine the location hint patterns for the rDNS names of these results (\eg \texttt{lis} in \texttt{server.lis50.r.cloudfront.net}).
We find locations for all destination addresses except for Google and YouTube where only \sperc{33} of names contain a usable location hint.

\begin{figure}
    \centering
    \includegraphics{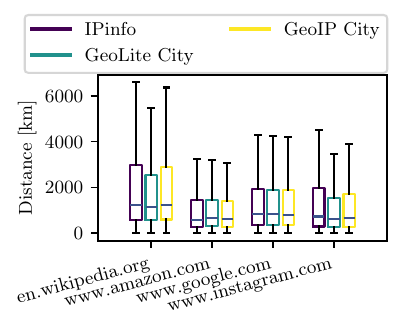}
    \vspace*{-25pt}
    \caption{Distance from the computed destination address location to the client subnet location obtained with the corresponding database.}%
    \label{fig:distances}
\end{figure}

We assign this location to clients based on \ecs{} results and compare the location to data from IPinfo~\cite{ipinfo}, and MaxMind's GeoLite City~\cite{maxmind_geolite} and GeoIP City databases~\cite{maxmind_geocity}.
\Cref{fig:distances} shows the distance between our estimate and the results from databases.
The boxplots show the median as a horizontal line and the quartiles Q1 and Q3 as box.
The whiskers are based on the 1.5 interquartile range value.
Distances to all databases are similar.
The median is between \SI{500}{\kilo\meter} and \SI{800}{\kilo\meter} for the three large operators and \SI{1200}{\kilo\meter} for Wikipedia.
A larger distance for Wikipedia is not surprising as it only had five active servers during the scanning time.
Nevertheless, the data shows that \sperc{75} of clients are within a radius of \SI{2000}{\kilo\meter} of their assigned \ac{pop} (\SI{3000}{\kilo\meter} for Wikipedia).

Large distances between the mapping of providers and the databases indicate that one information is inaccurate.
We select 25 client addresses where the distance between the database and the provider information for at least six domains is more than \SI{10000}{\kilo\meter}.
To estimate the correct geolocation we rely on distributed \ac{rtt} measurements with RIPE Atlas.
We select probes within the country indicated by the databases and providers.
Our results show that for 17 addresses, the \ac{rtt} from probes close to the database location is below \SI{40}{\milli\second} while probes close to the \ecs{} provider mapping result in \acp{rtt} larger \SI{100}{\milli\second}.
Thus, due to the physical limitations of transmission distance within a certain time, results indicate a better mapping by the databases.
For four addresses, results are inverted, indicating a better mapping from the \ecs{} provider.
Interestingly, probes close to both mappings for four targets result in low \acp{rtt}, indicating IP anycast deployments.

This evaluation shows, that for most client addresses, the \ecs{} provider mapping is similar to geolocation databases and the information can be used as an alternative.
In cases of large differences, both sources can be correct.
Further investigation is needed to improve this mapping or improve databases based on extracted \ecs{} provider information.
Our scanner reduces the number of queries to obtain the data relevant for this use case by up to \sperc{97}.

\subsection{TLS Scanning}
Scanning the \ac{tls} landscape is an important tool for researcher to learn about the deployment of new versions and extensions~\cite{amann2017diginotar,holzx5092011,holztls132020,kotzias2018tls,zirngibl2021quic} and to measure how many hosts are impacted by vulnerabilities~\cite{durumeric2014heartbleed}.
The \ac{sni} \ac{tls} extension allows a client to indicate the domain name of the requested service to the server during the \ac{tls} handshake.
For instances serving multiple domain names, this value is decisive to select the correct certificate.
\citeauthor{zirngibl2024libraryhunter}~\cite{zirngibl2024libraryhunter} report that scanning with \ac{sni} is especially important with QUIC~\cite{rfc9000} (which uses TLS 1.3) as some servers even drop the connection without an error when the Client Hello does not contain an \ac{sni} value.

Therefore, we evaluate the usefulness of our \emph{virtually distributed} \ac{ecs} results.
\Cref{tbl:fullscanstats} and \ref{tbl:approach_comp} show that our \ecs{} scans find significantly more addresses per domain than a simple DNS resolution from one vantage point.
For Google, we obtain \sk{2.2} IPv4 and \sk{2.1} IPv6 addresses while a single DNS response receives at maximum six different addresses, as it would be seen by a traditional single vantage point scan.
We compare the (domain, IP address)-pairs of our \ecs{} scans to our local DNS resolution of \sm{638} domain names from CZDS zone files~\cite{czds}, top lists, and domains in certificates from Certificate Transparency logs.
Out of the combined list of \sk{5.6} IP addresses (see \Cref{tbl:fullscanstats}) \sk{5.3} do not have any associated domain within our local DNS resolution.
A QUIC scan using the QScanner~\cite{zirngibl2021quic} reveals that we can successfully connect to \sk{4.6} of these targets when the server name is included in the \ac{sni}.
However, if we remove the \ac{sni} the scan is not successful for 300 IP addresses anymore but results in a timeout.
For the (domain, IP address)-pairs from our country prefix scan (see \Cref{ssec:moreprefixeval}) we find \sk{60} additional IP addresses paired with \sk{200} domains where we can complete a QUIC connection successfully.
For \sk{5.7} of these addresses, a handshake without \ac{sni} fails with an error and for \sk{35.6} addresses, the handshake even fails with a timeout.
Considering the \ac{sni} problems described by \citeauthor{zirngibl2024libraryhunter}~\cite{zirngibl2024libraryhunter}, we present this as a contribution to complement the existing measurement approach.

Even though TLS connections are not commonly dropped when no SNI extension is given, the server might not be able to return the appropriate certificate without it.
Our QUIC scan and additional \ac{tls} over TCP scans using the Goscanner~\cite{goscanner}, find two or more valid certificates per domain for all domains except the ones served by AWS Route53 (Amazon, TikTok, and PrimeVideo).
This evaluation already shows that \ecs scanning is beneficial for single vantage point \ac{tls} and QUIC scans to evaluate specific providers and certificates compared to a single vantage point.

\subsection{IPv6 Hitlist}%
\label{sec:ipv6hitlis}
The IPv6 Hitlist~\cite{gasser2018clusters, zirngibl2022rustyclusters} is an important measurement campaign that collects responsive IPv6 addresses.
While it is collecting data since 2018 from different sources, including different vantage points, most scans are conducted from a single vantage point.
Our \ecs{} scans allow us to evaluate the coverage of the hitlist based on our \emph{virtually distributed} scans and whether our \ecs{} approach provides further addresses.
In total, our scans reveal \sm{73.2} IPv6 addresses.
Interestingly, \sm{72.6} (\sperc{99.2}) of the addresses are not part of the cumulative input of the IPv6 Hitlist.
However, \sperc{99.8} of these addresses are within prefixes announced by Amazon.
According to the maintainers of the IPv6 Hitlist, Amazon is one of the main operators announcing so-called fully responsive prefixes~\cite{zirngibl2022rustyclusters}.
For those prefixes, each address is responsive to ICMP and port scans.
Thus, they are filtered from the hitlist service to reduce biases.
We compare newly found addresses to the list of fully responsive prefixes from the service and can verify that \sperc{94.7} are indeed within the already identified prefix and would be filtered.
We assume that Amazon encodes information (\eg{} related to the domain or resolution time) into the returned IPv6 address.
\citeauthor{zirngibl2022rustyclusters}~\cite{zirngibl2022rustyclusters} suggest that higher-layer scans should pay attention to fully responsive prefixes, but \emph{interesting} targets should be selected.
Our tool provides means to identify those targets based on domains and the load balancing strategy by the operators in addition to single query DNS resolutions.

Considering domains and results from the scans evaluated in \Cref{tbl:approach_comp} besides \texttt{www.amazon.com}, at least \sperc{85} of identified addresses for each domain are already known to the IPv6 Hitlist service.
Thus, their cumulative effort and combination of different sources covers most of the infrastructure of those well-known services.
Nevertheless, this work shows that \ecs{} and our tool can be used to identify the infrastructure of specific services in more detail.

\received{December 2024}
\received[revised]{April 2025}
\received[accepted]{April 2025}

\label{lastpage}

\end{document}